\documentclass{article}
\usepackage{graphicx} 
\usepackage{amssymb}
\usepackage{amsmath} 
\usepackage{biblatex}
\usepackage{hyperref}
\usepackage{algpseudocode}
\usepackage{algorithm}
\title{Upper and Lower Bounds on the Complexity of Unit-Distance Graph Realization in $\mathbb{Z}^d$}
\author{Eric Binnendyk}
\date{October 26, 2026}

\begin{document}

\maketitle

\begin{abstract}
    It is known that the problem of recognizing the existence of a unit-distance realization in $\mathbb{Z}^d$ for a graph $G$ is NP-complete. Here, assuming ETH, we give a sub-exponential lower bound of $2^{\Omega(\sqrt{n})}$ for this problem in $\mathbb{Z}^2$ and $2^{\Omega(n^{2/3})}$ in $\mathbb{Z}^3$, and give an upper bound of $2^{O(n^{(d-1)/d} (\log n)^2)}$ for the time complexity with an explicit divide-and-conquer algorithm. This approach beats naive exhaustive search by a factor of $\tilde{\Omega}(n^{1/d})$ in the exponent. For 2 and 3 dimensions, our upper and lower bounds match to within a logarithmic factor. 
\end{abstract}

\section{Introduction and related work}

The problem of \textbf{unit-distance graph recognition} for a metric space $S$ is to determine, for a graph $G = (V, E)$, whether there is an injective function $f: V \rightarrow S$ such that $\|f(v_1) - f(v_2)\|_2 = 1$ for every edge $(v_1, v_2)$. Such a function $f$ is a \textbf{unit-distance realization} of $G$ in $S$.

In this paper, we study the complexity of unit-distance graph recognition for $\mathbb{Z}^d$, the $d$-dimensional square grid lattice. This problem is also called \textbf{partial grid recognition} by de S\'{a} et al. \cite{desa2011} We call a unit-distance realization of in $\mathbb{Z}^d$ a $d$-dimensional \textbf{grid embedding} for short.

This problem is known to be NP-complete for $d = 2$ dimensions by a result of Bhatt and Cosmadakis, reducing the problem to NA3SAT (a variation of 3SAT) \cite{bhatt1987}, and is also NP-complete for $d > 2$ dimensions due to the graph prism construction introduced by de S\'{a} et al. \cite{desa2011}

One might wonder if we can determine anything about the exact complexity of this problem, assuming $P \ne NP$. A naive approach to this problem is to place one vertex at a specified location, do a depth-first search on $G$, and exhaustively try each of the $2d$ possible locations for each vertex that are one unit away from its parent. For each location, check that all cross-edges and back-edges are also unit distance. This algorithm runs in time $O((2d)^{|V|})$. One might wonder if, under suitable hardness assumptions, a $2^{\Omega(|V|)}$ algorithm is the best possible.

In this paper, we show that it is not, and we present a $2^{o(|V|)}$ algorithm for partial grid recognition based on the paradigm of divide-and-conquer across light balanced cuts of $G$. A cut of $G$ is defined to be \textbf{light} if it is crossed by fewer than $2n^{(d-1)/d}$ edges, and \textbf{balanced} if it has at least $cn$ vertices on both sides, where $c$ is some positive constant independent of $n$ (but dependent on $d$). The idea behind our sub-exponential search algorithm is that if we can prove the existence of a light balanced cut in any grid-embeddable graph $G$ that corresponds to a geometric cut in $\mathbb{Z}^d$, we can reduce the problem to finding such a cut and embedding the induced subgraphs of $G$ in non-overlapping bounding boxes on either side of the cut in such a way that the cut-adjacent vertices line up.

The problem of partial grid recognition is equivalent to determining if $G$ is an edge-induced subgraph of a grid graph, as defined in Wolfram MathWorld: \cite{mathworld}. As such, partial grid recognition is a special case of the subgraph isomorphism problem, which is known to be NP-complete in general, although there are known polynomial time algorithms when restricted to certain subclasses of graphs, such as planar graphs. \cite{eppstein1999}





Unit-distance graphs have been an object of mathematical study going back to the 1940s with work by Paul Erd\H{o}s. Erd\H{o}s posed the \textbf{unit distance problem} in 1946, which asks, given $n$ points in $\mathbb{R}^2$, what is the maximum number of pairs of points that can be one unit apart. A lower bound of $\Omega(n^{1 + \varepsilon})$, where $\varepsilon \approx 6.24 \times 10^{-38}$, was recently proved by an OpenAI model \cite{openai2026}, and was subsequently improved to $\varepsilon > 0.014$ by Will Sawin. \cite{sawin2026}

The problem of unit-distance graph detection has mainly been studied in $\mathbb{R}^2$, in which space it was shown by Schaefer to be $\exists \mathbb{R}$-complete by reduction to a geometric problem called Stretchability. \cite{schaefer2013} The class $\exists \mathbb{R}$ consists of all problems with polynomial-time reductions to the problem of deciding $\Sigma_1$ formulas over $\mathbb{R}$. To our knoweldge, no previous work determines the complexity of this problem in $\mathbb{R}^d$ for $d > 2$. At the end of \cite{schaefer2013}, Schaefer says that it is an open question to understand the complexity of decision problems involving linkages in higher-dimensional real space.

The basis of Bhatt and Cosmadakis' NP-hardness proof for unit-distance realizability in $\mathbb{Z}^2$ is the \textbf{logic engine} construction, simulating a logical formula with a graph. \cite{bhatt1987} Logic engines were also used by Eades and Whitesides in \cite{eades1996} to prove the hardness of the problem of nearest-neighbor realizability---the existence of a set of points in $\mathbb{R}^2$ such that the $|E|$ nearest pairs of points are precisely the edges of a graph $G$. A logic engine converts the satisfiability of an $n$-variable $m$-clause NA$k$SAT formula into the planar realizability of a graph with $O(mn)$ vertices and edges.

Other similar problems to unit-distance graph detection in $\mathbb{Z}^2$ have also been studied. For example, David Eppstein showed that if the restriction is added such that any path in $G$ of length $k$ has $L_1$ distance of $k$ in $\mathbb{Z}^2$, the recognition problem is in P. \cite{eppstein2005}

\section{Theorems about light balanced cuts}

These theorems are the basis of our sub-exponential search algorithm.

\subsection{Two-dimensional case}

First, we write out the proof for the two-dimensional case, which is simplest and contains the same ideas as the higher-dimensional proofs.

\textbf{Theorem 1}: For any 2-dimensional grid embedding of a graph $G$ with $n$ vertices, there is a horizontal or vertical cut passing through at most $2\sqrt{n}$ edges with at least $f(n) := (\sqrt{3} - 1)n - 2\sqrt{n}$ vertices on each side.

To prove this, we need some lemmas.

\textbf{Lemma 1}: Assume we have a set $V \subseteq \mathbb{Z}^2$ containing $n$ points where $pn$ points are in columns with at least $c\sqrt{n}$ points and $qn$ points are in rows with at least $c\sqrt{n}$ points, where $c > 1$. Then $q \le (1 - p)/(1 - p/c^2)$.

\textbf{Proof}: Call columns and rows with at least $c\sqrt{n}$ points \textbf{heavy}. Let $V_1 \subseteq V$ be the union of all heavy columns and $V_2 \subseteq V$ be the union of all heavy rows.

Let $C_1$ be the set of all heavy columns and $C_2$ be the set of all heavy rows. Because the elements of $C_1$ are pairwise disjoint with cardinality at least $c\sqrt{n}$, $|C_1| \le (p/c)\sqrt{n}$. Similarly, $|C_2| \le (q/c)\sqrt{n}$.

Every heavy row and heavy column intersect at exactly one point in $\mathbb{Z}^2$, which may or may not be a point in $V$. Every point in $V_1 \cap V_2$ is at one of these intersections. Therefore, $|V_1 \cap V_2| \le |C_1| \cdot |C_2| \le (pq/c^2)n$.

We can also find a lower bound on $|V_1 \cap V_2|$. We know that $|V - V_1| = (1 - p)n$ and $|V - V_2| = (1 - q)n$. Therefore, by a union bound, $|V - (V_1 \cap V_2)| \le (1 - p)n + (1 - q)n = (2 - p - q)n$, and so $|V_1 \cap V_2| \ge (p + q - 1)n$.

Putting our upper and lower bound together, we have:

\[ (p + q - 1)n \le (pq/c^2)n \]
\[ p + q - 1 \le pq/c^2 \]

Solving for q, we get:

\[ q(1 - p/c^2) \le 1 - p \]

Since $c > 1$ and $0 \le p \le 1$, we have $1 - p^2/c^2 > 0$. Therefore:

\[ q \le (1 - p)/(1 - p/c^2) \]

QED.


\textbf{Corollary}: Let $c = 2$ in the lemma above. Then $\min(p,q) \le 4 - 2\sqrt{3} \approx 0.536$.

\textbf{Lemma 2}: Call a column or row in $V$ \textbf{heavy} if it has at least $2\sqrt{n}$ vertices, and call it \textbf{light} otherwise, as above. There is a light column or row $C$ such that there are at least $f(n)$ points with smaller $x$-coordinate \textit{and} at least $f(n)$ points with larger $x$-coordinate (or $y$-coordinate respectively) than points in $C$, where $f(n) := (\sqrt{3} - 3/2)n - 2\sqrt{n}$.

\textbf{Proof of lemma 2}: Let $V \subseteq \mathbb{Z}^2$ be the set of vertex locations of $G$. As above, call a column or row in $V$ \textbf{heavy} if it has at least $2\sqrt{n}$ vertices. Let $V_1$ and $V_2$ be the sets of points in all heavy columns and all heavy rows respectively. By the above corollary, either $|V_1| \le (4 - 2\sqrt{3})n$ or $|V_2| \le (4 - 2\sqrt{3})n$. Without loss of generality, assume $|V_1| \le (4 - 2\sqrt{3})n$. In this case, we will prove the existence of a light column with $f(n)$ points on either side.

Sort the elements of $V = \{v_1, \dots, v_n\}$ by increasing $x$-coordinate, breaking ties arbitrarily. Let $w$ be the median point or one of the two median points, depending on if $n$ is odd or even. If $w$ is in a light column then we are done, because we know that at least $n/2 - 2\sqrt{n} > f(n)$ points are on either side of this column.

Otherwise, let $v_i$ be a point minimizing $\Delta := |(n + 1)/2 - i|$ over all points in light columns. Then the set $S = \{ v_j : |(n + 1)/2 - j| < \Delta \}$ is a subset of $V_1$. Therefore, $2\Delta - 1 = |S| \le |V_1| \le (4 - 2\sqrt{3})n$ or $\Delta \le (2 - \sqrt{3})n + 1/2$.

We know $i \ne (n + 1)/2$ because we assumed the column containing the median point(s) is heavy. Without loss of generality, assume $i < (n + 1)/2$. The same logic works when $i > (n + 1)/2$ with the sides reversed. Write $v_i = (x_i, y_i)$ and let $V' = \{(x, y) \in V: x < x_i\}$ and $V'' = \{(x, y) \in V: x > x_i\}$. Then $|V''| \ge n/2 > f(n)$. From $\Delta \le (2 - \sqrt{3})n + 1/2$ we get $i \ge (n + 1)/2 - (2 - \sqrt{3})n - 1/2 = (\sqrt{3} - 3/2)n$. Because $v_i$ is in a light column, $|V'| \ge i - 2\sqrt{n} \ge (\sqrt{3} - 3/2)n - 2\sqrt{n} = f(n)$. QED.

We now give the \textbf{proof of Theorem 1}. Let $V \subseteq \mathbb{Z}^2$ be the set of vertex locations of $G$. Lemma 2 says that there is either a column or a row of vertices $C$ with at least $f(n)$ vertices on each side. Without loss of generality, assume $C$ is a column.

Consider the set of edges $E'$ between points in $C$ and points in the next column $C'$, whose $x$-coordinates differ by 1. Because all edges go orthogonally, an edge in one column cannot connect to multiple edges in a neighboring column. Thus, $|E'| \le |C| \le 2\sqrt{n}$ and there are at least $f(n)$ vertices on each side of $E'$ including the vertices in $C \cup C'$. QED.

\subsection{Higher-dimensional case}

The proof of light balanced cuts in $d$ dimensions is very similar to the proof in 2 dimensions, so we skip the parts of the proofs that are almost identical.

The proof of correctness in 2 dimensions can be extended to $d$ dimensions as follows.

\textbf{Definition}. An \textbf{$i$-layer} is a subset of $V$ of the form $\{\mathbf{x} : x_i = k\}$.

\textbf{Lemma 1}: Let $V \subseteq \mathbb{Z}^d$ contain $n$ points. Let $V_i \subseteq V$ be the union of all $i$-layers with at least $cn^{(d - 1)/d}$ points (``heavy'' $i$-layers), where $c > 1$. Define $p_i := |V_i|/n$ for all $i$. Then there is a specific constant $p_{min} \in (0,1)$, dependent on $d$ and $c$, such that $\min_i p_i \le p_{min}$.

\textbf{Proof}: Let $C_i$ be the set of all heavy $i$-layers. Similar to the 2-dimensional case, we have $|C_i| \le (p_i/c) n^{1/d}$.

Also similar to the 2-dimensional case, we can find upper and lower bounds on the size of $\bigcap V_i$, the set of points in heavy layers in all directions. Every set of $d$ $i$-layers, with one for each direction $i$, intersect at exactly one point in $\mathbb{Z}^d$, which may or may not be a point in $V$. Therefore, $|\bigcap V_i| \le \prod_i ((p_i/c) n^{1/d}) = (\prod_i p_i)/c^d \cdot n$.

We can also find a lower bound on $|\bigcap V_i|$. We have $|V - V_i| = (1 - p_i) n$ and so $|V - \bigcap V_i| \le (d - \sum_i p_i) n$ and $|\bigcap V_i| \ge (\sum p_i - d + 1) n$.

As in the 2-dimensional case, the comparison between these bounds gives us a bound on a symmetric polynomial in all the $p_i$:

\[ (\sum_i p_i - d + 1)n \le (\prod_i p_i)/c^d \cdot n \]
\[ \sum_i p_i - d + 1 \le (\prod_i p_i)/c^d \]
\[ (\prod_i p_i)/c^d - \sum_i p_i + d - 1 \ge 0 \]

What condition on the $p_i$ does this imply? First, assume all $p_i$ are equal to a common value $p$, and let $P(p) = p^d/c^d - dp + d - 1$. Then the condition is equivalent to

\[ P(p) \ge 0 \]

We have $P(0) = d - 1 > 0$, $P(1) = 1/c^d - 1 < 0$ (because $c > 1$). We also have $P'(p) = dp^{d - 1}/c^d - d < d/c^d - d < 0$ for $0 < p < 1$. By the intermediate value theorem, there is exactly one root of $P$ in $(0,1)$. Call this root $p_{min}$. Over the range $(0,1)$, the condition $P(p) \ge 0$ is equivalent to $p \le p_{min}$.

Now go back to the original expression where the $p_i$ can be different. The partial derivatives of the expression with respect to each $p_i$ are negative for $p_i \in (0,1)$:

\[ \frac{\partial}{\partial p_i} \left({\frac{1}{c^d} \prod_j p_j - \sum_j p_j + (d - 1)}\right) = \frac{1}{c^d} \prod_{j \ne i} p_j - 1 \le \frac{1}{c^d} - 1 < 0 \]

This means we can replace each $p_i$ with a smaller value in $(0,1)$ without decreasing the value. In particular, we can replace all $p_i$ with $\min_i p_i$.

So our original inequality

\[ (\prod_i p_i)/c^d - \sum_i p_i + d - 1 \ge 0 \]

implies

\[ P(\min_i p_i) \ge 0 \]

which as we showed above is equivalent to

\[ \min_i p_i \le p_{min} \]

QED.

\textbf{Lemma 2}: Let $V \subseteq \mathbb{Z}^d$ contain $n$ points and let $c > 1$. Then there is a direction $i$ such that there is a particular $i$-layer $C \subseteq V$ defined by $C := \{\mathbf{x} \in V : x_i = k\}$, with $|C| < cn^{(d - 1)/d}$, where the sets $V^- = \{\mathbf{x} \in V : x_i < k\}$ and $V^+ = \{\mathbf{x} \in V : x_i > k\}$ both contain at least $(1 - p_{min})n/2 - cn^{(d - 1)/d}$ points.

\textbf{Proof}. As the proof is similar to what we had in the 2-dimensional case, we just sketch it. Let $i \in [d]$ be the direction given by \textbf{Lemma 1}. Sort all points in $V$ by $i$-coordinate. If the median point or pair of points are in a column with fewer than $cn^{(d - 1)/d}$ points, then $|V^-|, |V^+| > n/2 - cn^{(d - 1)/d}$ and we are done.

Otherwise, let $\mathbf{x}$ be the closest point to the median point or two points, in this ordering, that is in an $i$-layer with fewer than $cn^{(d - 1)/d}$ points; call this layer $C$. Let $S$ be the set of points closer to the median than $\mathbf{x}$; then $|S| \le p_{min} n$. The ranks of the points in $S$ range from $(n + 1)/2 - (|S| - 1)/2$ to $(n + 1)/2 + (|S| - 1)/2$. Assume $\mathbf{x}$ comes before the median point; then $\mathbf{x}$ has rank $(n + 1)/2 - (|S| - 1)/2 - 1 \ge (1 - p_{min})n/2$. We have $|C| < cn^{(d - 1)/d}$, so $|V^-| > (1 - p_{min})n/2 - cn^{(d - 1)/d}$ and $|V^+| > n/2 > (1 - p_{min})n/2 - cn^{(d - 1)/d}$. If $\mathbf{x}$ comes after the median point, the reasoning is similar. QED.

Note: If we knew $p_{min} < 1/2$, we could simplify the argument by only considering the last point in the lightest column before the median. That would require restricting the value of $c$, but we haven't yet tried this approach.

\textbf{Theorem 2}: For any $d$-dimensional grid embedding of a graph $G$ with $n$ vertices, there is a plane perpendicular to one of the axes passing through fewer than $2n^{(d - 1)/d}$ edges with at least $(1 - p_{min})n/2 - 2n^{(d - 1)/d}$ vertices on each side, where $p_{min} \in (0,1)$ is a constant that depends only on $d$. (Note that for any value of $d$, $(1 - p_{min})n/2 - 2n^{(d - 1)/d} > 0$ for sufficiently large $n$.)

\textbf{Proof}. Let $i$ be the direction and $C$ be the $i$-layer given by Lemma 2 with $c = 2$. Consider the set of edges $E'$ between points in $C$ and points in the next layer $C'$, whose $x$-coordinates differ by 1. Because all edges go orthogonally, an edge in one column cannot connect to multiple edges in a neighboring column. Thus, $|E'| \le |C| < 2n^{(d - 1)/d}$ and there are at least $(1 - p_{min})n/2 - 2n^{(d - 1)/d}$ vertices on each side of $E'$ including the vertices in $C \cup C'$. QED.

\section{Our realization algorithm}

Here, we present an exhaustive divide-and-conquer approach to determine partial grid recognition of a graph in time $2^{\tilde{O}(n^{(d - 1)/d})}$, beating naive search by a factor of $\tilde{\Omega}(n^{1/d})$ in the exponent, by making use of the above theorems about light balanced cuts.

Our algorithm makes use of a recursive function $GridEmbed(G, d, \mathbf{x}, pts)$. This function takes in a graph $G = (V, E)$, a dimension parameter $d$, a bounding box $\mathbf{x} = (x_1, \dots, x_d)$, and a function $pts : S \subseteq V \rightarrow \mathbb{Z}^d$ stored as a table mapping vertices to points. We either return an injective function $g: V \rightarrow \prod_i \{0, \dots, x_i\}$ such that for all $v \in S$, $g(v) = pts(v)$, and for each $\{v_1, v_2\} \in E$, $\|g(v_2) - g(v_1)\| = 1$, or determine that no such function exists. The function $g$ is also stored as a table mapping vertices to points.

We start with the base cases.

\subsection{Base cases}
One type of base case is a common situation for which we can guarantee no realization function exists. We immediately return $False$ on these cases:

\begin{itemize}
    \item If any $v \in V$ has degree $d(v) > 2d$.
    \item If any point in the range of $pts$ is outside the bounding box.
    \item If two inputs in $S$ map to the same output under $pts$.
    \item If the same input in $S$ maps to two outputs under $pts$ (i.e. $pts$ isn't actually a function). This could happen if the new cut was adjacent to a vertex that already had a mapping in the previous call.
\end{itemize}

We also have a base case for which the existence of a realization function is trivial. If $|V| < 2$ and $pts$ assigns a unique location inside the bounding box to the $0$ or $1$ points in $|V|$, we return $pts$.

\subsection{Main iteration and recursion}
We iterate through every position of an orthogonal planar cut within $\prod_i \{0, \dots, x_i\}$. Each is given by a direction $i$ and a pair of consecutive $i$-coordinates $j, j + 1$ where $0 \le j < x_i$. We also iterate through all $E' \subseteq E$ of cardinality $|E'| < 2n^{(d - 1)/d}$, and check if each $E'$ is a cut in $G$ with a graph search algorithm. If $E'$ contains two edges that share a vertex, skip this iteration (as it cannot be an orthogonal cut in a grid embedding). If $k$ is the number of connected components of $G$, each $E'$ is a cut for $2^k$ ordered pairs $(V', V'')$ partitioning $V$. We iterate through each such pair. We also iterate through every way to put the cut edges between the two hyperplanes: every injective function $f: E' \rightarrow \left(\prod_{i'} \{0, \dots, x_{i'}\}\right) \cap \{\mathbf{x}: x_i = j\}$ 

For each $i, j, E', V', V'', f$, we assume that the cut $(V', V'')$ maps to a planar cut through $\mathbb{Z}^n$ in direction $i$ and position $j$ with crossing edges in positions given by $f$. We make two recursive calls.

First, let $G'$ and $G''$ be the subgraphs induced by $V'$ and $V''$ respectively. Define new bounding boxes $\mathbf{x}_1 = (x_1, \dots, x_{i - 1}, j, x_{i + 1}, \dots, x_d)$ and $\mathbf{x}_2 =\\(x_1, \dots, x_{i - 1}, x_i - (j + 1), x_{i + 1}, \dots, x_d)$ by splitting the old bounding box in two. Define $V_j$ and $V_{j + 1}$ to be the vertices in $V'$ and $V''$ respectively that are adjacent to edges in $E'$. Define $pts_1: (S \cap V') \cup V_j \rightarrow \prod_i \{0, \dots, x_{1,i}\}$ by $pts_1(v) = pts(v)$ if $v \in V'$ and $pts_1(v) = f(e)$ for $v \in V_j$ where $e \in E'$ is the edge adjacent to $V_j$. Define $pts_2: (S \cap V'') \cup V_{j + 1} \rightarrow \prod_i \{0, \dots, x_{2,i}\}$ similarly, except where the vector $-(j + 1)e_i$ is added to each output.


Now make two calls $GridEmbed(G', d, \mathbf{x}_1, pts_1)$ and $GridEmbed(G'', d, \mathbf{x}_2, pts_2)$. If either call returns $False$, continue to the next loop iteration. Otherwise, both calls return embedding functions of their graphs, $g'$ and $g''$. Combine these functions into an embedding function $g: V \rightarrow \mathbb{Z}^d$ by defining $g(v) = g'(v)$ for $v \in V'$ and $g(v) = g''(v) + (j + 1)e_i$ for $v \in V''$ and return $g$.

\subsection{Ending condition}

If no iterations find valid realizations of $G'$ and $G''$, return $False$.

\subsection{The parent function call}

To determine partial grid recognition of a connected graph $G$ in $\mathbb{Z}^d$, we call\\$GridEmbed(G, d, (n - 1)^d, \varnothing)$, where $|V| = n$ and $(n - 1)^d$ is the length-$d$ vector whose entries are all $n - 1$. If $G$ is disconnected, we call the same function on all of its components and re-position the resulting embeddings so that they don't overlap.

\subsection{Proof of correctness}

Now we move on to the correctness proof of the algorithm, which follows from the theorems about light balanced cuts.

\textbf{Claim}: The function call $GridEmbed(G, d, \mathbf{x}, pts)$ returns a $d$-dimensional grid embedding of $G$ with bounding box extending from $(0)^d$ to $\mathbf{x}$ and with $v$ embedded at $pts(v)$ for $v$ in the domain of $pts$, if such an embedding exists, and returns \textit{false} otherwise.

\textbf{Proof of claim}: First we observe that the recursion always terminates, because the subgraphs in the recursive calls have strictly fewer vertices than the original graph. Therefore, we eventually reach the base cases of 0 and 1 vertices, for which correctness is easily verified.

Assume by strong induction that $GridEmbed$ is correct on all graphs with fewer than $n$ vertices. Consider a call $GridEmbed(G, d, \mathbf{x}, pts)$ where $G = (V, E)$ and $|V| = n$.

($\Rightarrow$:) First, assume that $G$ has a grid embedding $f: V \rightarrow \mathbb{Z}^d$ satisfying the above criteria. Theorem 2 says that there is a cut across fewer than $2n^{(d-1)/d}$ edges between two parallel layers of vertices where each side has at least $(1 - p_{min})n/2 - 2n^{(d - 1)/d}$ vertices. The rest of the embedding consists of two subgraphs $G'$ and $G''$ within smaller bounding boxes made by the intersection of the original bounding box with the two parallel planes.

The algorithm goes through all possible cuts $E'$ with $|E'| < 2n^{(d-1)/d}$, so it finds this cut. It also goes through all possible positions of cut edges across all parallel layers in the bounding box, so it eventually finds the locations of the cut edges in the embedding.

The restriction of $f$ to subgraphs $G'$ and $G''$ gives two grid embeddings of $G'$ and $G''$ in smaller bounding boxes. However, \textit{any} embeddings of $G'$ and $G''$ in their respective bounding boxes, as long as they have cut-edge-adjacent vertices in the same locations in addition to the restriction given by $pts$, can be connected with the cut edges $E'$ to get another grid embedding $f'$ of $G$. (Note that because the two bounding boxes don't overlap, the vertices in the embedding $f'$ won't overlap either.) We know that grid embeddings of the subgraphs exist. The algorithm $GridEmbed$ modifies $pts$ to add the restrictions about cut-edge-adjacent vertices in the two recursive calls. By the inductive hypothesis, some valid grid embeddings of $G'$ and $G''$ will be returned. Because they are inside the bounding boxes and have vertices in the designated places, they can be combined with $E'$ to make a valid embedding of $G$, which will be returned by the parent call.

($\Leftarrow$): Now assume that $G$ has no valid grid embedding respecting the restrictions. We showed above that the only possible realization that $GridEmbed$ can return is a valid grid embedding. Since there is no grid embedding of $G$, no realization can be returned, and so $GridEmbed$ must return $False$. QED

\subsection{Time complexity}

Let $T(n, d, x, k)$ be the time complexity of $GridEmbed$ on a graph with $n$ vertices embedded in $\mathbb{Z}^d$ with a maximum bounding box edge length of $x$ and in $k$ components.

Then we have the following upper bound for the time complexity (note that we have $|E| \le 2nd$ due to the preconditions): 

\[ T(n, d, x, k) \]
\[ \le dx(2nd)^{2n^{(d - 1)/d}}(x^{(d-1)/d})^{2n^{(d-1)/d}} 2^k \cdot 2T((\frac{1+p_{min}}{2}+\varepsilon)n, d, x, k + 2n^{(d-1)/d}) \]


for any $\varepsilon > 0$, for sufficiently large $n$ relative to $d$.

Specifically, we want $2n^{(d-1)/d} \le \varepsilon n$ so that $\frac{1-p_{min}}{2} \cdot n - 2n^{(d-1)/d} \ge (\frac{1-p_{min}}{2} - \varepsilon)n$. Then, $(\frac{1+p_{min}}{2}+\varepsilon)n$ is an upper bound on the number of vertices in each of $G'$ and $G''$.

Let $c := \frac{1+p_{min}}{2}+\varepsilon$. Note that $0 < c < 1$ for sufficiently small $\varepsilon$. Then we can unroll the recurrence relation into the following expression:

\[ T(n, d, x, k) \le \prod_{i = 0}^{O(\log n)} (dx(2nc^i d)^{2(nc^i)^{(d-1)/d}} (x^{(d-1)/d})^{2(nc^i)^{(d-1)/d}} 2^{k + 2in^{(d-1)/d} + 1}) \]

(Here, the exponent in the last term should contain a sum of $(nc^i)^{(d-1)/d}$ for different values of $i$, but these terms have been replaced by the larger value $n^{(d-1)/d}$ to make calculation easier.)

We can rewrite some of these exponential expressions with a base of 2:

\[ T(n, d, x, k) \le d^{O(\log n)} x^{O(\log n)} \left( \prod_{i = 0}^{O(\log n)}2^{2(nc^i)^{(d-1)/d} (1 + \log n + i \log c + \log d + (d-1)/d \cdot \log x)} \right) \]
\[ \cdot 2^{O(k \log n + (\log n)^2 n^{(d-1)/d})} \]

\[ \le (nd)^{O(\log n)} \cdot 2^{O(n^{(d-1)/d}(\log n + \log d + \log n \cdot (d - 1)/d + (\log n)^2) + k \log n)} \]

\[ \le 2^{O(n^{(d-1)/d}((\log n)^2 + \log d) + k \log n)} \]

Note that because $0 < c < 1$, $\sum_{i = 0}^{O(\log n)} c^i = O(1)$ and can be absorbed into the big $O$.

If $G$ starts with disconnected components on the outermost call, we can call the algorithm separately for each disconnected component. So we can assume without loss of generality that $k = 1$. In this case, because $\log n \ll n^{(d-1)/d}$, we have the following bound on the time complexity for a constant $d$:

\[ T'(n, d) \le 2^{O(n^{(d-1)/d} (\log n)^2)} \]

This beats exhaustive search by a factor of $O(n^{1/d} \log d/(\log n)^2)$ in the exponent.

\section{Lower bounds}

We now show how we can use NA$3$SAT reductions and the \textbf{Exponential Time Hypothesis} to find lower bounds of $2^{\Omega(\sqrt{|V|})}$ for unit-distance graph recognition in $\mathbb{Z}^2$ and $2^{\Omega(|V|^{2/3})}$ for the same problem in $\mathbb{Z}^3$.

First we show how to use Bhatt and Cosmadakis' reduction, along with the graph prism construction technique given in \cite{desa2011}, to produce a lower bound of $2^{\Omega(\sqrt{n})}$ ($2^{\Omega(\sqrt{n/2^d})}$ if not assuming constant $d$) for the $d$-dimensional grid embedding problem, assuming ETH. This nearly matches the upper bound when $d = 2$.

\subsection{Exponential Time Hypothesis and Sparsification\\Lemma}

\textit{Note: On the day this paper was published to Arxiv, OpenAI announced a claimed disproof of ETH by means of an algorithm for $k$-SAT in $2^{O(n/\sqrt{\log n})}$ time. This proof is currently under review. We expect that our lower bounds still hold with an extra factor of $n^\varepsilon$ in the exponent under a weaker version of ETH that asserts there is no algorithm with time $2^{O(n^{1-\varepsilon})}$ for 3-SAT instances with $O(n)$ clauses for any $\varepsilon > 0$, but this will need to be checked separately.}

The Exponential Time Hypothesis is a conjecture formulated by Russell Impagliazzo and Mohan Paturi, which is the basis of many lower bounds in fine-grained complexity. \cite{impagliazzopaturi1999}

\textbf{Definition}. The \textbf{Exponential Time Hypothesis} says that $k$-SAT cannot be solved in time $2^{o(n)}$ for each $k \ge 3$, where $n$ is the number of variables.


We will also need another tool to demonstrate that we can get away with making reductions from $n$-variable $k$-SAT instances with $O(n)$ clauses only. This tool is the \textbf{sparsification lemma}, introduced in \cite{impagliazzopaturizane1998} by Impagliazzo, Paturi, and Zane in 1998. It says that for any $k$-SAT formula there is a formula with a constant multiple of $n$ clauses with the same difficulty:

\textbf{Theorem - sparsification lemma}. There is an algorithm $A$ such that for every $k$-CNF formula $\Phi$, the call $A(k, \varepsilon, \Phi)$ runs in $poly(n) 2^{\varepsilon n}$ time and outputs a sequence $\Psi_1, \dots, \Psi_t$ of $k$-CNF formulae, where $t \le 2^{\varepsilon n}$, each $\Psi_i$ has $O(n)$ clauses, and the solution set of $\Phi$ is the union of the solution set of each $\Psi_i$.

\textbf{Corollary (Calabro, Impagliazzo, Paturi)} \cite{calabroimpagliazzopaturi2006}. Let SAT$_\Delta$ be the problem of SAT restricted to input formulae with at most $\Delta \cdot n$ clauses. For each $k$ and $\varepsilon > 0$, there is a value $\Delta$ that is $poly(1/\varepsilon)$ (for constant $k$) giving an $O(2^{\varepsilon n})$ time Turing reduction from $k$-SAT to SAT$_\Delta$ instances with the same number of variables.

Finally, because we are dealing with NA$k$SAT instead of $k$-SAT, we will need to demonstrate a reduction from $k$-SAT to NA$k$SAT for each value of $k$. We do this by applying \textbf{clause splitting} to a standard reduction from $k$-SAT to NA$(k+1)$SAT.

\textbf{Theorem}. Let $m$ be any function of $n$. If $k$-SAT with $n$ variables and $m$ clauses requires $2^{\Omega(n)}$ time, then NA$k$SAT with $n + m + 1$ variables and $2m$ clauses requires $2^{\Omega(n)}$ time.

\textbf{Proof}. Let $\varphi = \bigwedge_{i=1}^m C_i$ be a $k$-CNF formula with $n$ variables. For each $C_i = (\ell_{i,1}, \dots, \ell_{i,k})$, we define $C_i' = (\ell_{i,1}, \dots, \ell_{i,k}, s)$, where $s$ is a dummy variable. We define the NA$(k+1)$SAT formula $\psi = (C_1', \dots, C_m')$. Then, any satisfying assignment for $\varphi$ becomes a satisfying assignment for $\psi$ with $s = 0$. Furthermore, every satisfying assignment for $\psi$ can be assumed to have $s = 0$, possibly by negating variables. Then it corresponds to a satisfying assignment for $\varphi$. Now we define a NA$k$SAT formula $\psi' = (C_{1,1}', C_{1,2}', \dots, C_{m,1}', C_{m,2}')$ by a technique called clause splitting. We introduce new variables $z_1, \dots, z_m$ and define $C_{i,1}' = (\ell_{i,1}, \dots, \ell_{i,k-1}, z_i)$ and $C_{i,2}' = (\ell_{i,3}, \dots, \ell_{i,k}, s, \bar{z_i})$. If $\psi$ has a satisfying assignment, it can be converted to a satisfying assignment for $\psi'$ by assigning $z_i = \neg \bigwedge \ell_{i,1}, \dots, \ell_{i,k-1}$ for each $i$. Furthermore, if $\psi'$ has a satisfying assignment, the conditions on $z_i$ imply that the subset of values assigned to variables in $\psi$ produces a satisfying assignment for $\psi$. The formula $\psi'$ has $n + m + 1$ variables and $2m$ clauses. If a sub-$2^{\Omega(n)}$ algorithm for NA$k$SAT exists with these parameters, then the reduction can construct a sub-$2^{\Omega(n)}$ algorithm for $k$-SAT. QED.

\subsection{A $2^{\Omega(|V|^{2/3})}$ lower bound for 3 dimensions}

Here, we describe how to build a reduction from $n$-variable NA3SAT to 3D grid embedding of a graph with $O(n^{3/2})$ vertices. This shows that, assuming ETH, the problem of 3D grid graph embedding has time complexity $2^{\Omega(|V|^{2/3})}$.

Our construction uses a 3D version of the logic engine that encodes each variable as a set of connections between lines perpendicular to the center plane, taking advantage of the third dimension to avoid overlaps. There are $O(n)$ such lines in total, and we use a graph coloring argument to show that $O(\sqrt{n})$ layers are required for all the connections.

\subsubsection{The base}

\begin{figure}
    \centering
    \includegraphics{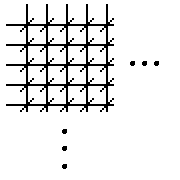}
    \caption{A corner of the 3D grid base. The ``diagonal'' lines go in the $\pm z$ directions.}
\end{figure}

The base of the 3D logic engine consists of a grid of vertices with two extra edges coming out of each vertex, three coming out of each edge vertex, and four coming out of each corner vertex. Each of these edges connects a grid vertex $a$ to a ``tassel'' vertex $a'$ that connects to no other vertices in the base.

This grid has only one embedding, besides rigid transformation and the rearrangement of non-grid edges. To see this, name the grid vertices $a_{i,j}$ according to their coordinate, where $i$ and $j$ range from $1$ to some value that is $O(n)$. Each square of grid vertices must be embedded as a square in $\mathbb{Z}^3$. Consider a single square $A$ with vertices $a_{i,j}, a_{i,j+1}, a_{i+1,j}, a_{i+1,j+1}$. Assume that $A$ is embedded horizontally with a $z$-coordinate of 0.

Note that each vertex of $A$ has six neighbors. These six neighbors must take up all six adjacent locations in the grid. In a given embedding, let $a'$ and $a''$ be the two tassel vertices adjacent to a grid vertex $a$ and embedded above and below the plane $z = 0$.

Consider edges $(a_{i,j}, a_{i,j+1})$ and $(a_{i,j+1}, a_{i,j+2})$. They must be parallel in the embedding. That is because $(a_{i,j+1}, a_{i,j+2})$ must be perpendicular to $(a_{i,j+1}, a_{i+1,j+1})$, and in two of the three options, $a_{i,j}'$ or $a_{i,j}''$ would overlap with one of the six vertices adjacent to $a_{i,j+2}$. Similarly, edges $(a_{i,j}, a_{i+1,j})$ and $(a_{i+1,j}, a_{i+2,j})$ must be parallel. If two edges of a grid square have fixed embeddings at a right angle, the other two edges also have a unique embedding in the same plane. Starting from vertex $a_{0,0}$, it follows by induction that the entire grid is embedded in a plane in $\mathbb{Z}^3$.

\subsubsection{Chains}

\begin{figure}
    \centering
    \includegraphics{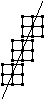}
    \caption{Part of a chain}
\end{figure}

A \textbf{chain} is a simple path with extra edges added to ensure the path is straight. Each chain $c$ is made of vertices $e_k$ for $k = 1, \dots, h$, where $h$ is the height. A $3 \times 3$ grid graph of points $L_k$ (that is, a $2 \times 2$ grid graph of squares), called a link, is connected to each vertex $e_k$, with $e_k$ being identified with the middle point in the grid.

Two chains are added to the tassels around each grid point $a_{4i-2,4j-2}$, where $i$ and $j$ range from $1$ to $O(\sqrt{n})$, spanning the entire grid. These chains are indexed as $c_{i,j}$ and $c_{i,j}'$, and are called the \textbf{positive chain} and \textbf{negative chain} respectively. They are connected to the grid as follows: To $a_{4i-2,4j-2}'$ we add an edge to vertex $e_1$ of $c_{i,j}$ and to $a_{4i-2,4j-2}''$ we add an edge to vertex $e_1$ of $c_{i,j}'$.

We can prove that, in this setup, all but the last two links of any chain must be embedded in horizontal planes parallel to the base:

\textbf{Theorem}. Let $G$ be a graph consisting of a base and a chain $e_1, \dots, e_{h+2}$ connected by an edge $(a’, e_1)$, where $a’$ is a tassel vertex attached to an internal grid vertex. In any grid embedding of $G$, the first $h$ links $L_1, \dots, L_h$ must be in $h$ consecutive planes parallel to the base with $e_1, \dots, e_h$ in a line perpendicular to the base.

\textbf{Proof}. We proceed by strong induction. Assume that for $0 \le k < h$, the first $k$ links are embedded in parallel planes. Assume without loss of generality that $a’$ is a tassel pointing upwards from the base in the $+z$ direction. If $k = 0$, then $e_{k+1}$ must be in the plane above $a’$ because all other directions are taken by tassels or the base. If $k > 0$, $e_{k+1}$ must be in the plane above $L_k$ for the same reason. Of the four neighbors of $e_{k+1}$ in $L_{k+1}$, assume by contradiction that one of them $v$ is above $e_{k+1}$. Without loss of generality, say that $v$ points in the $+z$ direction instead of the $+x$ direction. Then $e_{k+2}$ has to be in the $+x$ direction from $e_{k+1}$. The vertex $e_{k+2}$ has six neighbors, including $e_{k+3}$, but it is one unit away from a non-neighbor in the $-z$ direction: either a tassel vertex (if $k = 0$) or a vertex in $L_k$ (if $k > 0$). This implies that there is not enough room to embed all neighbors of $e_{k+2}$, a contradiction. Therefore, $L_{k+1}$ can only be embedded as a link above and parallel to link $k$. QED.

Note that neighboring chains are four units apart and the embedding of each chain consisting of parallel links takes up a region of size 3 by 3 by $O(n)$. This is small enough that embeddings of multiple chains do not overlap. Therefore, the part of the graph consisting of the base and the chains has a number of 3-dimensional grid embeddings where each chain in a pair can be flipped up or down, perpendicular to the base. Apart from the embedding of the last link, each link must be embedded parallel to the base.

\begin{figure}
    \centering
    \includegraphics{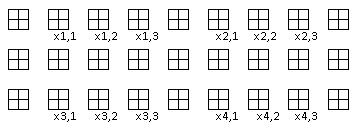}
    \caption{A single layer of chain links associated with variables in a 4-clause formula. Note that any two of the variables could be identical, but they still get separate chains.}
\end{figure}

\subsubsection{Flags}

\begin{figure}
    \centering
    \includegraphics{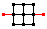}
    \caption{A link with two flags added on opposite sides}
\end{figure}

Flags are a tool we will use to encode satisfiability constraints, as in the original logic engine design. For a chain link with center vertex $e$, a \textbf{flag} is a vertex $f$ connected to one of the four neighbors $e'$ of $e$. Observe that if there are links above and below $e$, or if $e$ is the closest link to the base, $f$ must be on the same layer as $e$, but it can be rotated or reflected into any of eight configurations (of which at most four are distinct up to labeling of vertices). The only way a chain link with flags could have no embedding is if, no matter how it is positioned, the flag overlaps with a flag of a different chain. Because neighboring chains are four units apart, this can only happen if two neighboring chains have flags pointing towards each other.

\subsubsection{Encoding NA3SAT constraints}

\begin{figure}
    \centering
    \includegraphics[width=0.8\linewidth]{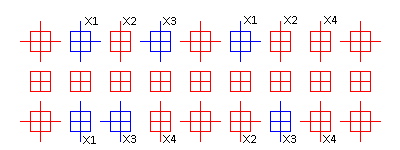}
    \caption{The layer directly above the base in the encoding of the NA3SAT formula $\varphi = (X_1, X_2, X_3), (X_1, X_2, X_4), (\neg X_1, X_3, X_4), (\neg X_2, \neg X_3, \neg X_4)$. The embedding uses the satisfying assignment $X_1 = X_3 = 0, X_2 = X_4 = 1$. Red chains mean the positive side is pointing up, blue chains mean the negative side is pointing up.}
\end{figure}

Let $\varphi$ be the NA3SAT formula, with $m$ clauses. By the sparsification lemma, we can assume $m \in O(n)$.

Make sure the base of the graph is sufficiently large to accommodate all gadgets, but still $O(\sqrt{n})$ by $O(\sqrt{n})$.

Encode each clause in $\varphi$ as a sequence of three consecutive horizontal chains with at least one empty row/column of chains in the horizontal/vertical directions between clauses.

To encode the clauses, add flags to the two links of each chain closest to the base. For the literal $u_{i,j}$ in clause $C_i$, let $c$ be the corresponding chain. Let $e_{+}$ and $e_{-}$ be the nearest links on the positive and negative sides. If $u_{i,j}$ appears positively in $C_i$, add two flags to opposite sides of $e_{+}$ and three flags on three sides of $e_{-}$. If $u_{i,j}$ appears negatively, do the same with $e_{-}$ and $e_{+}$ swapped.

Put flags on all four sides of a link that is between two clauses in the same row.

Leave off all flags from the links in rows between clauses.

\textbf{Theorem}. Let $\varphi = (C_1, \dots, C_m)$ be a NA3SAT formula and let $G$ be a logic engine graph with chains and flags added to encode $\varphi$ (but no connections added; see below). For any family of assignments $f_i: \{x_{i,1}, x_{i,2}, x_{i,3}\} \rightarrow \{True, False\}$, the following two statements are equivalent:
\begin{itemize}
    \item Each $f_i$ is a satisfying assignment for $C_i$.
    \item There is an embedding of $G$ such that the chain corresponding to $x_{i,j}$ points in the $+z$ direction if $f_i(x_{i,j})$ is true and the $-z$ direction otherwise.
\end{itemize}

\textbf{Proof}. We focus on the three chains for the variables in $\varphi$ and the surrounding chains. The two neighboring chains in each row have four flags on them, so no matter how they are oriented, a flag is sticking towards the first chain from the left and the third chain from the right. If the first chain has three flags, one of the flags must be pointing to the right as well. The same applies to the second chain, and so on by induction. If all chains have three flags, the third chain must have a flag pointing to the right, which clashes with the chain to the right. Thus no embedding is possible. On the other hand, if at least one link $e$ has two flags, there is an embedding where every link in the row has flags pointing forward and backward, every link before $e$ has flags pointing to the right, and every link after $e$ has flags pointing to the left. The links on neighboring rows have no flags and so they do not interfere with this construction.

Furthermore, without adding connections, there is always a way to embed the chains corresponding to each clause (just set two literals to true and one to false, for instance). So without any other restrictions, an embedding of a single clause can be extended into an embedding of the rest of the structure.

So an embedding exists if and only if, on both the layers above and below the base, one of the links in $\varphi$ has two flags instead of three. For the layer above the base, this means there is either a positive literal with a true variable or a negative literal with a false variable; for the layer below the base, vice versa. Thus, the constraints of NA3SAT are enforced. QED.


\subsubsection{Encoding variable connections}

We have a way of encoding satisfiability of individual clauses. But we need a way of ensuring that the same variable in different clauses gets the same truth value. We accomplish this with variable connections.

\textbf{Definition}. Between any two chains $c_{i_1,j_1}$ and $c_{i_2,j_2}$, a \textbf{variable connection} at height $h$ is a simple path between link $2h$ of the positive side of chain $c_{i_1,j_1}$ and link $2h + 1$ of the positive side of chain $c_{i_2,j_2}$, connected to vertices which are themselves adjacent to the center of the links, and consisting of $2 + 4(i_2 - i_1 + j_2 - j_1)$ new vertices if $j_2 \ge j_1$ and $-2 + 4(i_2 - i_1 + j_2 - j_1)$ new vertices if $j_2 < j_1$, unless $i_1 = i_2$ in which case there are $4|j_2 - j_1|$ new vertices.

Variable connections are not allowed to be placed on the first link (because of the flags) or the last two links of a chain (which are not guaranteed to be horizontal).

\begin{figure}
    \centering
    \includegraphics[width=0.8\linewidth]{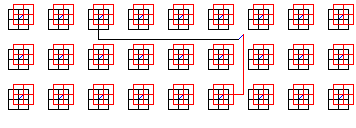}
    \caption{A variable connection between two chains (lower layer = black, upper layer = red)}
\end{figure}

\textbf{Canonical embeddings}. Any variable connection has a standard embedding that is possible if the two chains either both point up or both point down. This embedding consists of starting from the link of $c_{i_1,j_1}$ (where we assume that $j_2 > j_1$), going one step in the positive $y$ direction, going left or right until the $x$-coordinate matches the empty space to the right of column $j_2$, going up in the $z$ direction one step, and then going in the positive $y$ direction until we are directly to the right of the link of $c_{i_2,j_2}$, and then going one step to the left to connect. Exception: If $i_2 = j_2$, we simply go forward one unit in the $y$ direction, to the right, and then backward one unit to connect to the second link. It can be shown that this is always possible by rotating the links into the right orientation because of the number of vertices in the path.

We now show that canonical embeddings allow a variable connection to be embeddable precisely when it should be.

\textbf{Theorem}. Consider a graph $G$ with a base, a field of chains, and flags on upper and lower layer 1. For any valid embedding $f$ of $G$ without connections added and any $h \ge 1$, $f$ can be extended to an embedding with a connection between chains $c_1$ and $c_2$ at height $h$, if and only if the positive sides of $c_1$ and $c_2$ either both point up or both point down.

\textbf{Proof}. If they do point in the same direction, the endpoint links are on the same side. We can use the ``standard'' embedding described above on any two layers above layer 1, and there will be no clashes because the two layers have no flags.

If they do not point in the same direction, first note that the Manhattan distance between the two points in the canonical embedding is at most two units less than the length $\ell$ of the path in edges. Rotating the endpoints on the chains decreases the Manhattan distance by at most two on each side, so the minimum possible Manhattan distance between the endpoints when they are on the same side is $\ell - 6$. The Manhattan distance between the $xy$-components of the points is at least $\ell - 7$. If one of the two chains is flipped to the other side, the $z$-coordinates are at least $6$ units apart. Therefore, the Manhattan distance between the two endpoints when they are on opposite sides is at least $\ell - 1$ units. Therefore, any embedding of the connection between endpoints on opposite sides cannot go in two opposite directions at different steps, as this will require the path length to be at least the Manhattan distance plus two. Therefore, every point of the path must be inside the bounding box between the endpoints. But every such point on the base layer is taken by a vertex of the base, so there will be a clash. Thus, no embedding exists when the links are on opposite sides. QED.

However, we will need to add multiple variable connections to a single graph. We need a theorem telling us when two variable connections will never interfere with each other's embeddability.

\textbf{Definition}. The variable connection between $c_1$ and $c_2$ \textbf{overlaps} with the connection between $c_3$ and $c_4$ if any two chains, one from each pair, have the same $x$ or $y$ coordinate.

\textbf{Theorem}. If variable connections are added to the graph in the above theorem, the variable connection from $c_1$ to $c_2$ at height $h$ is embeddable if $c_1$ and $c_2$ point in the same direction \textit{and} the VC does not overlap with any other variable connection at height $h$. The VC from $c_1$ to $c_2$ is not embeddable if $c_1$ and $c_2$ point in opposite directions.

\textbf{Proof}. Notice that in the standard embedding, all the new vertices are placed on two lines: the row whose $y$-coordinate is two more than the center of chain $c_1$ on layer $2h$ (above or below the base), and the column whose $x$-coordinate is two more than the center of chain $c_2$ on layer $2h + 1$. If two VCs on the same layer clash, then they must either use the same horizontal or vertical line for their vertices. Thus, they overlap in the sense described above. Furthermore, the ``lower'' layer of one connection cannot clash with the ``upper'' layer of another, because the former has even index and the latter has odd index. Thus if the value $h$ is different for two connections, the two connections don't overlap. QED.

To show that we can fit all our variable connections on a single graph without clashing, we need a special theorem, where we model the chains as a set of 2D grid points:

\textbf{Theorem}. Given an $m$ by $n$ grid of vertices with connections between certain pairs of points where each point has at most $d$ connections incident to it. Draw each connection as a horizontal segment followed by a vertical segment (either of these segments can be of length 0). If the vertical parts or the horizontal parts of any two connections are in the same column or row respectively, they need to be colored different colors. Then there exists a coloring of all connections with at most $(n + m)d - 1$ colors and there is an efficient algorithm to find it.

The proof of this theorem uses an ingredient called Brooks' Theorem, which gives an upper bound on the chromatic number of a graph with bounded degree. Brooks' theorem is a classic theorem of graph theory, first proven by R. Leonard Brooks in 1941. \cite{brooks1941} The intuition behind Brooks' Theorem is that if you build a graph, at some point you have so many vertex colors that a new vertex can't conflict with all of them.

\textbf{Theorem} (Brooks' Theorem). For any undirected graph $G$ with maximum vertex degree $\Delta$, there is an efficient greedy algorithm to find a coloring of the vertices of $G$ with $\Delta + 1$ colors, where no two neighboring vertices have the same color. In fact, if the graph is not a complete graph or a cycle, $\Delta$ colors are needed.

Because we're dealing with asymptotic complexity, we don't need the conditional bound of $\Delta$ colors and can use the greedy coloring bound of $\Delta + 1$ colors instead. The proof of the greedy bound is simply that, if $\Delta + 1$ colors are used, there will be at most $\Delta$ colors that are unusable for each vertex, so the first color will be at most index $\Delta + 1$.

From Brooks' Theorem, we can prove our theorem about grid connection coloring by modeling the grid connections as a graph.

\textbf{Proof of above theorem}. Build a graph $G$ whose vertices are the grid connections, and edges are pairs of connections. Due to the maximum number of connections of each vertex, each connection overlaps (is in the same column/row as) at most $(n + m)d - 2$ other connections. Thus, the max degree of $G$ is at most $\Delta = (n + m)d - 2$. By Brooks' Theorem, there is an efficient coloring of the vertices of $G$ with at most $(n + m)d - 1$ colors, which leads to a coloring of connections where no two connections with the same color share the same row/column.

We now use the above theorems about grid connections on an abstract level to prove a bound on the number of layers needed for a certain number of connections.

\textbf{Theorem}. If a 3D logic engine graph has an $m$ by $n$ grid of chains and each chain has at most $d$ connections, there is an algorithm to place the connections in $2(n + m)d - 2$ layers so that they don't clash.

\textbf{Proof}. We build an abstract $m$ by $n$ grid where each point represents a chain and connect each pair of points that require a connection. Now the above theorem on colored connections applies. The condition for two connections to get different colors in the theorem (horizontal or vertical components in the same line) exactly matches the condition for two variable connections to get different layers to avoid clashing. This implies that, if we run the coloring algorithm, we can put connections of each color on a separate pair of layers and none will clash. The algorithm yields a solution with $(n + m)d - 1$ colors. Because each connection needs two layers, we use $2(n + m)d - 2$ layers. QED.

\subsubsection{Proof of main construction}

\textbf{Main theorem - 3-dimensional partial grid recognition reduction}. There is a polynomial time reduction from any $n$-variable $O(n)$-clause NA3SAT formula $\varphi$ to a graph $G$ with $O(n^{3/2})$ vertices that has a unit-distance embedding in $\mathbb{Z}^3$ if and only if $\varphi$ is satisfiable.

\textbf{Proof}. Build the base and the chains as described above, in an $O(\sqrt{n})$ by $O(\sqrt{n})$ grid, with the chains representing variables in each clause of $\varphi$ and with extra chains between neighboring clauses. Add flags on the link layers nearest the base to encode the local NA3SAT constraints. For each Boolean variable $X_i$ in $\varphi$, list all the chains $X_{i,1}$, $X_{i,2}$, etc. corresponding to variable $X_i$. Put a connection between each $X_{i,j}$ and $X_{i,j+1}$. This ensures that in any embedding, chains $X_{i,j}$ and $X_{i,j+1}$ have the same truth value. By extension, all chains representing the same variable have the same truth value. Furthermore, the maximum number of connections each chain participates in is $d = 2$. Thus, the length of each chain is $O(\sqrt{n})$. This also ensures the maximum number of connections is $O(n)$.

The base of the graph has $O(n)$ vertices. There are $O(n)$ chains and each chain has $O(\sqrt{n})$ vertices. There are also $O(n)$ connections each with $O(\sqrt{n})$ vertices. Thus, the entire graph has $O(n^{3/2})$ vertices. QED.

\subsection{A $2^{\Omega(|V|^{2/3})}$ lower bound for 3-dimensional grid embedding}

Assuming ETH, the reduction from NA3SAT to 3-dimensional grid embedding implies a lower bound for the complexity of the latter problem.

\textbf{Theorem}. Assuming ETH, 3-dimensional grid embedding takes time $2^{\Omega(|V|^{2/3})}$.

\textbf{Proof}. We use the sparsification lemma corollary with $k = 3$. Assuming ETH, on some input with $n$ variables, any algorithm $A$ for 3-SAT takes at least $2^{cn}$ for some $c$. Choose any $0 < \varepsilon < c$, and use the corollary to fix a constant $\Delta$.

Let $\varphi$ be an instance of 3-SAT with $n$ variables. We use the corollary to reduce $\varphi$ to an instance of SAT$_\Delta$ with $n$ variables and $O(n)$ clauses. Then we use the reduction to convert this instance of SAT$_\Delta$ to NA3SAT also with $O(n)$ clauses, and then to a graph $G$ using our new reduction for grid embedding in $\mathbb{Z}^3$. Then we check if $G$ is has a 3-dimensional grid embedding. The first reduction takes $O(2^{\varepsilon n})$ and makes $O(2^{\varepsilon n})$ oracle calls. The second reduction (to NA3SAT and then to $G$) takes $poly(n)$ time, which means the algorithm for verifying grid realizability of $G$ must take $\Omega(2^{(c - \varepsilon)n})$ time. The graph $G$ has $O(m^{3/2})$ vertices, and $m \in O(n)$. Thus, for all $n$, $G$ has $|V| \in O(n^{3/2})$ vertices. The time complexity of the realizability algorithm is then lower bounded by $\Omega(2^{(c - \varepsilon) |V|^{2/3}}) \subseteq 2^{\Omega(|V|^{2/3})}$. This leads us to our conclusion: conditional on ETH, any algorithm for 3-dimensional grid embedding takes at least $2^{\Omega(|V|^{2/3})}$ runtime for at least one graph for each of an infinite number of values of $|V|$. This is sufficient to prove that $2^{\Omega(|V|^{2/3})}$ is a conditional lower bound for the complexity of this problem. 

QED.

\subsection{The graph prism construction: a $2^{\Omega((|V|/2^d)^{2/3})}$ lower bound for $d$-dimensional grid embedding}

In \cite{desa2011}, de S\'{a} et al. define the \textbf{prism} of a graph to be the result of connecting the matching vertices of two copies of the graph. We use this construction to extend our result to a conditional lower bound for $d$-dimensional grid embedding for all $d > 3$.

\textbf{Definition}. Let $G$ be $(V, E)$. Then the \textbf{prism} of $G$, $prism(G)$, is defined as $(\{v_i, v’_i : v_i \in V\}, \{(v_i, v_j), (v’_i, v’_j) : (v_i, v_j) \in E\} \cup \{(v_i, v’_i) : v_i \in V\})$. (This is also the Cartesian product of $G$ with the complete graph $K_2$.)

\begin{figure}
    \centering
    \includegraphics[width=7cm]{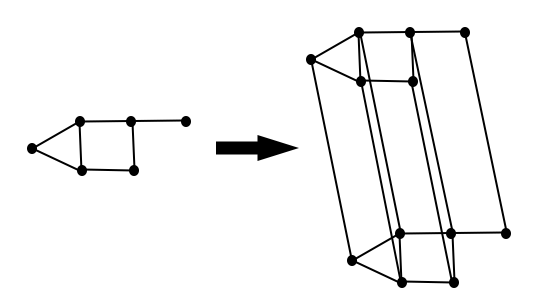}
    \caption{Example graph prism construction.}
\end{figure}

Note that the prism of a graph with $n$ vertices has $2n$ vertices.

Now we show that any $d$-dimensional grid embedding of $G$ leads to a $d+1$-dimensional grid embedding of $prism(G)$ and vice versa. The proof of the 2D $\rightarrow$ 3D case was given in \cite{desa2011}; our proof is a simple modification to arbitrary dimensions.

\textbf{Definition}. A \textbf{lateral edge} of $prism(G)$ is an edge $(v_i, v_i')$ between duplicate vertices in the two copies of $G$.

\textbf{Theorem}. If $G$ is connected, any $d$-dimensional grid embedding of $prism(G)$ must have the two copies of $G$ in neighboring parallel planes.

\textbf{Proof}: Let $f$ be a grid embedding of $prism(G)$. We proceed by structural induction on the vertices of $G$. For the base case, pick a vertex $v$ of $G$, and let $(v, v')$ be the corresponding edge of $prism(G)$. Let $i$ be the direction of the vector $f(v') - f(v)$. Now let $V'$ be a subset of vertices of $G$ containing $v$. Assume by induction that the lateral edges of each vertex in $V'$ are realized in direction $i$ between two parallel $i$-planes one unit apart. Let $v_2$ be a vertex in $G$ not in $V'$ that is connected to a vertex $v_1$ in $V'$. Let the corresponding lateral edges be $(v_2, v_2')$ and $(v_1, v_1')$ respectively. The only ways to realize $v_2$ and $v_2'$ are:
\begin{itemize}
    \item With $f(v_2) - f(v_1)$ being a vector in direction $i$. Then $f(v_2') - f(v_1')$ must be the same vector in order for $v_2$ and $v_2'$ to be $1$ unit apart. But this means that either $f(v_1) = f(v_2')$ or $f(v_1') = f(v_2)$, which are both impossible because vertices overlap. So this cannot happen.
    \item With $f(v_2) - f(v_1)$ and $f(v_2') - f(v_1')$ having $i$-coordinates of $0$. Again, these two vectors must be identical. In this case, the edge $(v_2, v_2')$ is parallel to $(v_1, v_1')$ and only changes on a coordinate not equal to $i$. So $(v_2, v_2')$ lies between the same parallel planes as the lateral edges corresponding to vertices in $V$.
\end{itemize}
Because $G$ is connected, this structural induction eventually reaches all vertices of $G$. Thus, all vertices in the same subgraph as $v_1$ have $i$-coordinates matching $v_1$, and the same with $v_2$. The entire realization therefore consists of two copies of $G$ is parallel $i$-planes. QED.

\textbf{Theorem}. A graph $G$ can be realized in $\mathbb{Z}^d$ if and only if $prism(G)$ can be realized in $\mathbb{Z}^{d + 1}$.

\textbf{Proof}.
\begin{itemize}
    \item $\Rightarrow$
\end{itemize}
If $G$ has a $d$-dimensional grid embedding, let $f$ be the realization function. Now define a realization function $f'$ for $prism(G)$ by $f'(v) = f(v) + (0)$, $f'(v') = f(v) + (1)$. It can be shown that no vertices are overlapping and every edge has unit length in realization $f'$.

\begin{itemize}
    \item $\Leftarrow$
\end{itemize}
Assume $prism(G)$ can be realized in $\mathbb{Z}^{d + 1}$. If $G_i$ is a connected component of $G$, then $prism(G_i)$ is a connected component of $prism(G)$. Thus, every connected component of $G$ is realized in two parallel planes. Each of these connected components can be realized separately in $\mathbb{Z}^d$ to give a grid embedding of $G$. QED.

\textbf{Theorem}. Assuming ETH, $d$-dimensional grid embedding takes time $2^{\Omega((|V|/2^d)^{2/3})}$.

\textbf{Proof}. Just like above, we assume ETH and reduce satisfiability with $n$ variables to 3-dimensional partial grid recognition of a graph $G$. The former problem, and hence the latter problem, must take $2^{\Omega(n)}$ time. We reduce to verifying grid realizibility in $\mathbb{Z}^d$ of the graph $prism^{d-3}(G)$, which has $|V| = O(n^{3/2} 2^d)$ vertices. Now our lower bound for solving realizability in $\mathbb{Z}^d$ is $2^{\Omega((|V|/2^d)^{2/3})}$. QED.

\section{Future research}

In the future, we plan to adapt this research to show similar lower bounds for the complexity of grid embedding in $d > 3$ dimensions. We conjecture that there is a reduction from $m$-variable NA3SAT to $d$-dimensional partial grid recognition of a graph with $O(m^{d/(d-1)})$ vertices, implying a conditional lower bound of $2^{\Omega(|V|^{(d-1)/d})}$ time. This matches our algorithm to a logarithmic factor in the exponent.

We also conjecture that a lower bound of $2^{\Omega(|V|^{(d-1)/d})}$ holds for the complexity of partial grid recognition on any discrete $d$-dimensional lattice with a basis of unit vectors, and that this lower bound is tight up to a logarithmic factor in the exponent.

We expect similar logic-engine-based constructions to work in these cases to prove NP-hardness, but it might be a bit harder to come up with a building block that we can ensure makes rigid structures.

We also want to research the hardness of unit-distance graph recognition in higher dimensions, not just in lattices but in $\mathbb{R}^d$. We expect that Schaeffer's Peaucellier-linkage based construction, which was used to show that unit-distance graph recognition in $\mathbb{R}^2$ is $\exists \mathbb{R}$-hard, can be extended to higher dimensions in a systematic way to show that the problem for $\mathbb{R}^d$ is also $\exists \mathbb{R}$-hard, but we are currently unaware of a specific extension.

\section{Acknowledgements and AI usage}

I give thanks to my academic advisors, Russell Impagliazzo and Mohan Paturi, for providing some critical insights regarding this paper. In particular, Russell suggested the idea of using a divide-and-conquer approach to show that partial grid recognition has a $2^{o(n)}$ time algorithm. He also gave me the idea of proving the existence of light near-balanced cuts with union bounds, and helped me understand the sparsification lemma. Mohan gave me feedback on readability and connections to other results. I also give thanks to the mathematician David Eppstein at UC Irvine for looking at an earlier draft of this paper and pointing out that some of my results were previously known.

I did not use generative AI to write any text in this paper or prove any results. However, I used Google’s AI to get the statement of Brooks’ theorem after I rediscovered the greedy graph coloring bound myself and wanted a citation, and I used ChatGPT for proofreading an earlier draft of this paper. ChatGPT helped me find and fix errors in the definition of the cut-edge placement function in the algorithm and the proof of the $k$-SAT $\rightarrow$ NA$k$SAT reduction.

\printbibliography

@inproceedings{schaefer2013,
  title={Realizability of Graphs and Linkages},
  author={Marcus Schaefer},
  year={2013},
  url={https://api.semanticscholar.org/CorpusID:8477664},
  booktitle={unknown}
}

@article{eppstein2005,
title = {The lattice dimension of a graph},
journal = {European Journal of Combinatorics},
volume = {26},
number = {5},
pages = {585-592},
year = {2005},
issn = {0195-6698},
doi = {https://doi.org/10.1016/j.ejc.2004.05.001},
url = {https://www.sciencedirect.com/science/article/pii/S0195669804000885},
author = {David Eppstein}
}

@INPROCEEDINGS{impagliazzopaturi1999,
  author={Impagliazzo, R. and Paturi, R.},
  booktitle={Proceedings. Fourteenth Annual IEEE Conference on Computational Complexity (Formerly: Structure in Complexity Theory Conference) (Cat.No.99CB36317)}, 
  title={Complexity of k-SAT}, 
  year={1999},
  volume={},
  number={},
  pages={237-240},
  doi={10.1109/CCC.1999.766282}}

@INPROCEEDINGS{impagliazzopaturizane1998,
  author={Impagliazzo, R. and Paturi, R. and Zane, F.},
  booktitle={Proceedings 39th Annual Symposium on Foundations of Computer Science (Cat. No.98CB36280)}, 
  title={Which problems have strongly exponential complexity?}, 
  year={1998},
  volume={},
  number={},
  pages={653-662},
  doi={10.1109/SFCS.1998.743516}}

@INPROCEEDINGS{calabroimpagliazzopaturi2006,
  author={Calabro, C. and Impagliazzo, R. and Paturi, R.},
  booktitle={21st Annual IEEE Conference on Computational Complexity (CCC'06)}, 
  title={A duality between clause width and clause density for SAT}, 
  year={2006},
  volume={},
  number={},
  pages={7 pp.-260},
  doi={10.1109/CCC.2006.6}}

@online{mathworld,
    title = "GridGraph",
    publisher = "Wolfram MathWorld",
    url = "https://mathworld.wolfram.com/GridGraph.html"
}

@article{eppstein1999,
  author       = {David Eppstein},
  title        = {Subgraph Isomorphism in Planar Graphs and Related Problems},
  journal      = {CoRR},
  volume       = {cs.DS/9911003},
  year         = {1999},
  url          = {https://arxiv.org/abs/cs/9911003},
  bibsource    = {dblp computer science bibliography, https://dblp.org}
}

@article{eades1996,
title = {The logic engine and the realization problem for nearest neighbor graphs},
journal = {Theoretical Computer Science},
volume = {169},
number = {1},
pages = {23-37},
year = {1996},
issn = {0304-3975},
doi = {https://doi.org/10.1016/S0304-3975(97)84223-5},
url = {https://www.sciencedirect.com/science/article/pii/S0304397597842235},
author = {Peter Eades and Sue Whitesides}
}

@article{bhatt1987,
title = {The complexity of minimizing wire lengths in VLSI layouts},
journal = {Information Processing Letters},
volume = {25},
number = {4},
pages = {263-267},
year = {1987},
issn = {0020-0190},
doi = {https://doi.org/10.1016/0020-0190(87)90173-6},
url = {https://www.sciencedirect.com/science/article/pii/0020019087901736},
author = {Sandeep N. Bhatt and Stavros S. Cosmadakis}
}

@article{desa2011,
title = {Complexity dichotomy on partial grid recognition},
journal = {Theoretical Computer Science},
volume = {412},
number = {22},
pages = {2370-2379},
year = {2011},
issn = {0304-3975},
doi = {https://doi.org/10.1016/j.tcs.2011.01.018},
url = {https://www.sciencedirect.com/science/article/pii/S0304397511000521},
author = {Vinícius G.P. {de Sá} and Guilherme D. {da Fonseca} and Raphael C.S. Machado and Celina M.H. {de Figueiredo}}
}

@article{openai2026,
title = {Planar Point Sets with Many Unit Distances},
author = {OpenAI},
url = {https://cdn.openai.com/pdf/74c24085-19b0-4534-9c90-465b8e29ad73/unit-distance-proof.pdf}
}

@misc{sawin2026,
      title={An explicit lower bound for the unit distance problem}, 
      author={Will Sawin},
      year={2026},
      eprint={2605.20579},
      archivePrefix={arXiv},
      primaryClass={math.CO},
      url={https://arxiv.org/abs/2605.20579}, 
}

@article{brooks1941, title={On colouring the nodes of a network}, volume={37}, DOI={10.1017/S030500410002168X}, number={2}, journal={Mathematical Proceedings of the Cambridge Philosophical Society}, author={Brooks, R. L.}, year={1941}, pages={194–197}}

\end{document}